\documentclass[10pt,twocolumn,aps,prl,nobibnotes,showpacs,superscriptaddress,floatfix]{revtex4-1}

\usepackage{graphicx}
\usepackage{tikz}
\usepackage{amssymb}
\usepackage{amsmath}
\usepackage{color}
\usepackage{psfrag}
\usepackage{epsfig}
\usepackage{bbm}
\usepackage{bm}
\usepackage{hyperref}
\usepackage[normalem]{ulem}
\hypersetup{breaklinks}
\usepackage{amsthm}

\newcommand{\beq}{\begin{eqnarray}}
\newcommand{\eeq}{\end{eqnarray}}

\newcommand{\A}[1]{\mathcal A}

\definecolor{mygreen}{rgb}{0,0.5,0}
\definecolor{myblue}{rgb}{0,0,0.75}
\definecolor{mymagenta}{cmyk}{0,1,0,0.12}

\usepackage{braket}
\usepackage{bbm}
\theoremstyle{plain}
\newtheorem{theorem}{Theorem}

\theoremstyle{definition}
\newtheorem{observation}[theorem]{\textbf{Observation}}

\theoremstyle{remark}

\DeclareMathOperator{\tr}{tr}

\begin{document}
\title{Unified picture for spatial, temporal and channel steering}

\author{Roope Uola}
\affiliation{Universit\"{a}t Siegen, Walter-Flex-Str. 3, D-57068 Siegen, Germany}
\author{Fabiano Lever}
\affiliation{Universit\"{a}t Siegen, Walter-Flex-Str. 3, D-57068 Siegen, Germany}
\affiliation{Universit\"{a}t Potsdam, Karl-Liebknecht-Str. 24, 14476 Potsdam
, Germany}
\author{Otfried G\"{u}hne}
\affiliation{Universit\"{a}t Siegen, Walter-Flex-Str. 3, D-57068 Siegen, Germany}
\author{Juha-Pekka Pellonp\"{a}\"{a}}
\affiliation{Turku Centre for Quantum Physics, University of Turku, FI-20014 Turku, Finland}

\begin{abstract}
Quantum steering describes how local actions on a quantum system can affect another, space-like separated, quantum state. Lately, quantum steering has been formulated also for time-like scenarios and for quantum channels. We approach all the three scenarios as one using tools from Stinespring dilations of quantum channels. By applying our technique we link all three steering problems one-to-one with the incompatibility of quantum measurements, a result formerly known only for spatial steering. We exploit this connection by showing how measurement uncertainty relations can be used as tight steering inequalities for all three scenarios. Moreover, we show that certain notions of temporal and spatial steering are fully equivalent and prove a hierarchy between temporal steering and macrorealistic hidden variable models.
\end{abstract}


\maketitle


\textit{Introduction.---} Quantum steering refers to the possibility of one party, typically called Alice, to affect the quantum state of a spatially separated party, typically called Bob, by making only local measurements on her system. Quantum steering formalises spooky action at a distance \cite{letter} and as such it is an entanglement verification method intermediate to trust-based entanglement witnesses and no trust-requiring device-independent scenarios, e.g., Bell inequalities. Steering provides a natural framework for semi-device independent quantum information protocols \cite{QKD, Piani1, Xiang} and a guideline for theoretical and experimental work on both entanglement theory and non-locality \cite{Toby, Tamas, Bowles, Wiseman, Loopholefree}. Moreover, steering is known to be closely connected to incompatibility of quantum measurements \cite{Tulio, Roope1}. To be more precise, it has been shown that steering and joint measurability problems are in one-to-one correspondence \cite{Roope2} and that unsteerability of quantum states can be checked through incompatibility breaking properties of quantum channels \cite{Jukka}.

\begin{figure}[h]
\centering
\begin{tikzpicture}
    \node at (0,0) {$\rho_{AB}$};
    \draw[thick, ->] (0.5,0) -- (2,0) node[right] {$\rho_{a|x}$};
    \draw[thick, ->] (-0.5,0) -- (-2.8,0);
    \draw (-3.4,-0.3) rectangle (-2.8,0.3) node[pos=0.5] {$A_x$};
    \draw[-]  (-3.1, -0.3) -- (-3.1, -1.15);
    \draw[-]  (-2.96, -0.18) -- (-2.96, -0.85);
    \draw[->] (-3.1, -1.15) -- (2,-1.15) node[pos=0.17, below] {\scriptsize $a$} node[right]{$a$} ;
    \draw[->] (-2.96, -0.85) -- (2,-0.85)  node[pos=0.15, above] {\scriptsize $x$} node[right]{$x$};

    \draw[-] (-3.6,-1.6) rectangle (-1,0.6);
    \draw[-] (1.5,-1.6) rectangle (2.8,0.6);
    \node at (-2.3, 0.8) {Alice};
    \node at (2.1, 0.8) {Bob};
\end{tikzpicture}
\begin{tikzpicture}
    \node at (-1.25,0) {$\rho$};
    \draw[thick, ->] (-1,0) -- (0,0);
    \draw (0,-0.3) rectangle (0.6,0.3) node[pos=0.5] {$\mathcal{I}_x$};
    \draw[thick, ->] (0.6,0) -- (2.7,0) node[right] {$\mathcal I_{a|x}(\rho)$};
    \draw[-]  (0.3, -0.3) -- (0.3, -1.15);
    \draw[-]  (0.44, -0.18) -- (0.44, -0.85);
    \draw[->] (0.3, -1.15) -- (2.7,-1.15) node[pos=0.17, below] {\scriptsize $a$} node[right]{$a$} ;
    \draw[->] (0.44, -0.85) -- (2.7,-0.85)  node[pos=0.15, above] {\scriptsize $x$} node[right]{$x$};

    \draw[-] (-0.35,-1.6) rectangle (1.05,0.6);
    \draw[-] (2.2,-1.6) rectangle (4,0.6);
    \node at (0.35, 0.8) {Alice};
    \node at (3.1, 0.8) {Bob};
\end{tikzpicture}
\caption{Spatial steering (top): Alice and Bob share a bipartite state $\rho_{AB}$, Alice measures $A_x$ and classically communicates the measurement setting ($x$) and result ($a$) to Bob. The (non-normalised) post-measurement state assemblage Bob receives is given as $\rho_{a|x}=\text{tr}_A[(A_{a|x}\otimes\openone)\rho_{AB}]$. Temporal steering (bottom): Alice applies an instrument $\mathcal{I}_x$ on a single system state $\rho$ and classically communicates the measurement setting ($x$) and result ($a$) to Bob, together with the (non-normalised) output state  $\mathcal I_{a|x}(\rho)$.}
\label{Fig1}
\end{figure}
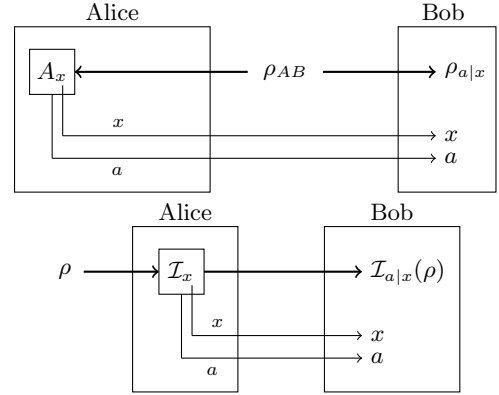

Extending the spatial case, steering has recently found its temporal counterpart \cite{YNChen} (see Fig.~\ref{Fig1}). The idea of temporal steering is to ask whether steering-like phenomenon can be achieved on a single quantum system, where Alice measures a single particle first and then hands it to Bob. Naturally one could argue that some sort of steering effect is easy to reach in such scenarios, because Alice's measurement choice can in principle affect Bob's state, i.e. Alice can signal to Bob. However, signalling can be excluded by using well-chosen input states. The remaining non-trivial (as well as trivial) scenarios have found connections to, for example, non-Markovianity \cite{SLChen}. In this work we want to characterise quantum measurements. Hence, we don't wish to restrict ourselves to specific input states. Instead, we take an approach where non-signalling is a state-independent built-in feature. Later we show that all (non-trivial) temporal scenarios can be mapped into our formulation.

State steering has a natural extension to the level of quantum channels through the well-known state-channel isomorphism \cite{Piani2}. This extension is called channel steering, and it investigates the possibility of Alice to affect Bob's end of a broadcast channel from Charlie to Alice and Bob. Technically, channel steering can be seen as a (semi-device independent) method to verify that an extension of a channel from Charlie to Bob is coherent \cite{Piani2}.

By now channel steering has been introduced as a theoretical construction, but in this article we show how a certain modification of the channel steering protocol can be used as a powerful framework for all three steering scenarios. To demonstrate the strength of our approach, we map all three steering scenarios one-to-one to the incompatibility of quantum measurements, provide universally applicable steering inequalities through measurement uncertainty relations, show an equivalence between spatial and temporal steering, and finally show a hierarchy between temporal steering and macrorealistic hidden variable models.

\textit{Spatial steering.---} Steering scenarios can be seen as processes where an untrusted party (Alice) sends a trusted party (Bob) a state assemblage $\{\rho_{a|x}\}_{a,x}$, where $x$ labels the measurements and $a$ the respective outcomes, satisfying the non-signalling condition $\sum_{a}\rho_{a|x}=\sum_{a}\rho_{a|x'}$ for all $x,x'$. The non-signalling property is crucial in our scenarios for reasons to become clear in the following subsections. The steerability of a state assemblage is decided by checking the existence of a so-called local hidden state model (see also below).

In {\it spatial} steering the state assemblage originates from space-like separated local measurements on one party and is hence naturally non-signalling. Formally, consider a bipartite system described by a quantum state $\rho_{AB}$. When Alice performs measurements described by positive operator valued measures (POVMs) $\{A_{a|x}\}_{a,x}$ (i.e.\ $A_{a|x}\geq0$ and $\sum_a A_{a|x}=\openone$ for all $a,x$)  on her system, Bob is left with a non-normalised state assemblage
\begin{align}
\rho_{a|x}:=\text{tr}_A[(A_{a|x}\otimes\openone)\rho_{AB}].
\end{align}
Here, $\openone$ is the identity operator on Bob's system. The setup is called (spatially) unsteerable if Bob can recover his state assemblage from a local state ensemble (or state preparator) $\{p(\lambda),\sigma_\lambda\}_\lambda$ together with additional information about Alice's choice of measurement $x$ and obtained outcome $a$ by means of classical post-processing, i.e.\ if for every $a,x$
\begin{align}\label{steerdef}
\rho_{a|x}=\sum_\lambda p(\lambda)p(a|x,\lambda)\sigma_\lambda
\end{align}
and steerable otherwise. Here $p(\lambda)\ge 0$ is the probability that Bob's state $\sigma_\lambda$ occurs and
$p(a|x,\lambda)\geq0$ are conditional probabilities so that $\sum_a p(a|x,\lambda)=1$ for each $x,\,\lambda$. The r.h.s.\ of Eq.~(\ref{steerdef}) is called, when existing, a local hidden state (LHS) model for the assemblage $\{\rho_{a|x}\}_{a,x}$.

\textit{Temporal steering.---} For temporal steering one needs the concept of quantum instruments. Quantum instruments are collections of completely positive maps which sum up to a completely positive trace preserving (cptp) map, i.e.\ to a quantum channel. Physically, instruments describe the state transformation caused by a measurement and one can think of them as a generalisation of the projection postulate to the case of POVMs. For a POVM $\{A_a\}_a$ the most typical instrument is the von Neumann-L\"uders instrument $\mathcal I^{L}_a(\rho)=\sqrt{A_a}\rho\sqrt{A_a}$ and all possible instruments compatible with $\{A_a\}_a$ are the ones which code the measurement outcome probabilities into the post-measurement state, i.e.\ $\text{tr}[\mathcal I_a(\rho)]=\text{tr}[A_a\rho]$ for all $\rho$.
It can be shown \cite{JP} that any such instrument implementing $\{A_a\}_a$ can be described by the quantum channels $\{\Lambda_a\}_a$ from Alice to Bob applied to the von Neumann-L\"uders instrument via
$
\mathcal I_a(\rho)=\Lambda_a\big[\mathcal I^{L}_a(\rho)\big].
$

In temporal steering one is interested in state assemblages $\{\rho_{a|x}^{\text{temp}}\}_{a,x}$ which are given by the actions of a set of quantum instruments $\{\mathcal I_{a|x}\}_{a,x}$ on a single system state $\rho_A$. The steerability of this assemblage is decided by checking the existence of a LHS model, i.e.\ the scenario is temporally non-steerable if
\begin{align}\label{tempsteerdef}
\rho_{a|x}^{\text{temp}}:=\mathcal I_{a|x}(\rho_A)=\sum_\lambda p(\lambda)p(a|x,\lambda)\sigma_\lambda
\end{align}
and steerable otherwise. In temporal steering the non-signalling condition is not a built-in feature. 
However, as some input states lead to steering trivially, it makes sense to talk about temporal steering only in the case of non-signalling assemblages. Finally, note that sometimes temporal state assemblages are defined through an instrument and an additional time evolution. However, as a concatenation of an instrument and a channel is an instrument, we don't write the channel explicitly to our state assemblages.

\textit{Main technique.---} As our main technique we use the Stinespring dilation of quantum channels. In text book quantum mechanics any quantum channel $\Lambda$ on a finite-dimensional system is given through the representation
\begin{align}\label{Stineuni}
\Lambda(\rho)=\text{tr}_E[U\rho_0\otimes\rho U^\dagger],
\end{align}
where $U$ is a unitary operator on the total space of the system and an environment $E$, and $\rho_0$ is a quantum state of the environment \cite{Stine}. This type of representation is, however, not the only way to dilate a channel. It appears that a slightly modified version of Stinespring dilation is better tailored for our purposes. Namely, instead of using a unitary operator on the system and its environment, we define an isometry $V:\mathcal H\rightarrow\mathcal A\otimes\mathcal K$, where $\mathcal H$ and $\mathcal K$ are the Hilbert spaces of the input and output systems and $\mathcal A$ is the Hilbert space of a dummy system. For a channel given in the Kraus form $\Lambda(\rho)=\sum_{k=1}^r K_k\rho K_k^\dagger$, the isometry $V$ can be constructed as $V|\psi\rangle=\sum_{k=1}^r|\varphi_k\rangle\otimes K_k|\psi\rangle$ for all $|\psi\rangle$ with $\{|\varphi_k\rangle\}_{k=1}^r$ being an orthonormal basis of the dummy system. With this isometry the dilation simply reads
\begin{align}\label{StineV}
\Lambda(\rho)=\text{tr}_{\mathcal A}[V\rho V^\dagger].
\end{align}
Note that this dilation doesn't have a specific initial state on the environment and, hence, in order to make a clear distinction between the text-book dilation in Eq.~(\ref{Stineuni}) and our dilation in Eq.~(\ref{StineV}) we talk about a dummy system instead of an environment.

We are specifically interested in sets of instruments $\{\mathcal I_{a|x}\}_{a,x}$ which do not allow signalling, i.e.\ which have the same total channel $\Lambda:=\sum_a\mathcal I_{a|x}=\sum_a\mathcal I_{a|x'}$ for every $x,x'$. Non-signalling instruments are related to observables on the dummy space of their total channel $\Lambda$ \cite{Werner, Teiko}. Namely, the actions of non-signalling instruments $\{\mathcal I_{a|x}\}_{a,x}$ can always be written as actions of a set of POVMs $\{\tilde A_{a|x}\}_{a,x}$ on the dummy part of the dilation, i.e.
\begin{align}\label{Stineins}
\mathcal I_{a|x}(\rho)=\text{tr}_{\mathcal A}[(\tilde A_{a|x}\otimes\openone)V\rho V^\dagger].
\end{align}
Note that in general the dummy POVMs $\{\tilde A_{a|x}\}_{a,x}$ do not coincide with the POVMs $\{A_{a|x}\}_{a,x}$ one measures on the actual system. Note, moreover, that the isometry $V$ is constructed from $\Lambda$ and does, due to non-signalling, not depende on $x$.

In what follows, we will mainly concentrate on minimal dummy systems, i.e. minimal Stinespring dilations. The minimality means that $r$ is the smallest possible dimension, in which case the Kraus operators of the total channel are linearly independent. In this case (for a given total channel) the correspondence between the dummy POVMs and the instruments they define is one-to-one \cite{Werner, Teiko}. Namely, we have that
\begin{align}\label{Bobsolve}
\mathcal I_{a|x}(\rho)=\sum_{k,l=1}^r\langle\varphi_l|\tilde A_{a|x}|\varphi_k\rangle K_k\rho K_l^\dagger
\end{align}
from where the matrix elements $\langle\varphi_l|\tilde A_{a|x}|\varphi_k\rangle$ of the dummy POVMs can be computed.

\textit{Minimal dilation for a state assemblage.---} A crucial concept for our study is joint measurability. Joint measurability of a set of POVMs $\{A_{a|x}\}_{a,x}$ is defined as the existence of a common POVM $\{G_\lambda\}_\lambda$ from which the original POVMs can be post-processed. This means that the set is jointly measurable if for every $a,x$
\begin{align}
A_{a|x}=\sum_\lambda p(a|x,\lambda)G_\lambda
\end{align}
and incompatible otherwise. Here $p(\cdot|x,\lambda)$ is a probability distribution for every $x,\lambda$.

Because of the one-to-one connection between dummy POVMs and instruments (for a given total channel), we see that a set $\{\tilde A_{a|x}\}_{a,x}$ of dummy POVMs is jointly measurable if and only if the instruments $\{\mathcal I_{a|x}\}_{a,x}$ (defined through the minimal dilation) have a common refinement, i.e. for every $a,x$ one has
\begin{align}\label{instrcomp}
\mathcal I_{a|x}=\sum_\lambda p(a|x,\lambda)\mathcal I_\lambda.
\end{align}

To relate this connection to steering, note that any state ensemble $\{p(\lambda),\rho_{\lambda}\}_\lambda$, where $\sum_\lambda p(\lambda)=1$ is an output of a state preparator, i.e.\ an instrument with a trivial input space $\mathbb C$. Even though using a one-dimensional Hilbert space might sound unconventional, it appears to be useful for our purposes as any non-signalling state assemblage $\{\rho_{a|x}\}_{a,x}$ can be seen as originating from a non-signalling set $\{\mathcal I_{a|x}\}_{a,x}$ of state preparators. Hence, any non-signalling state assemblage can be written [see Eq.~(\ref{Stineins})] through the minimal Stinespring dilation as
\begin{align}\label{non-signallingss}
\rho_{a|x}=\mathcal I_{a|x}(|1\rangle\langle 1|)=\tr_{\mathcal A}[(\tilde A_{a|x}\otimes\openone)|\psi\rangle\langle\psi|],
\end{align}
where $|1\rangle$ is a complex number with norm one and $|\psi\rangle:=V|1\rangle$ is a unit vector on the compound system. As the dummy observables $\{\tilde A_{a|x}\}_{a,x}$ are unique for a given minimal dilation, and as the state preparator corresponding to a LHS model has the same total channel as state preparators associated to the assemblage, we arrive to our first Observation $[$see also Eq.~(\ref{instrcomp})$]$

\begin{observation}\label{steerJM}
Any non-signalling state assemblage $\{\rho_{a|x}\}_{a,x}$ is unsteerable if and only if the associated observables $\{\tilde A_{a|x}\}_{a,x}$ on the minimal dilation of the corresponding state preparator are jointly measurable.
\end{observation}

In order to make the Observation~\ref{steerJM} more concrete, consider a state assemblage given by a set of state preparators $\{\mathcal I_{a|x}\}_{a,x}$ through Eq.~(\ref{non-signallingss}). The state of the total system $V|1\rangle\langle 1|V^\dagger$ is clearly a purification of $\rho_B:=\sum_a\rho_{a|x}$. One possible choice of this purification is the canonical one $|\psi\rangle=\mathbb (\mathbb I\otimes\rho_B^{1/2})|\psi^+\rangle$, where $|\psi^+\rangle=\sum_i|ii\rangle$ is the non-normalised maximally entangled state written in the eigenbasis of $\rho_B$. For this choice Eq.~(\ref{non-signallingss}) reads
\begin{align}
\rho_{a|x}&=\tr_{\mathcal A}[(\tilde A_{a|x}\otimes\rho_B^{1/2})|\psi^+\rangle\langle\psi^+|(\openone\otimes\rho_B^{1/2})]\nonumber \\
&=\rho_B^{1/2}\tilde A_{a|x}^T\rho_B^{1/2}\label{AliceBob},
\end{align}
where the transpose is taken in the eigenbasis of $\rho_B$. In this case the dummy observables whose joint measurability solves spatial and temporal steerability are given as $\tilde A_{a|x}=\rho_B^{-1/2}\rho_{a|x}^T\rho_B^{-1/2}$.

Noting that joint measurability is invariant under transposition, we get as a special case of our Observation~\ref{steerJM} the known result \cite{Roope2} for spatial steering stating that a state assemblage $\{\rho_{a|x}\}_{a,x}$ is unsteerable if and only if the so-called Bob's steering equivalent observables defined as $B_{a|x}:=\rho_B^{-1/2}\rho_{a|x}\rho_B^{-1/2}$ are jointly measurable.

It is worth mentioning that Observation~\ref{steerJM} can also be used to reproduce a known example of the connection between temporal steering and joint measurability for scenarios using L\"uders instruments and a maximally mixed input state \cite{Karthik}. For discussion about this example and its connection to Observation~\ref{steerJM}, see Appendix A.

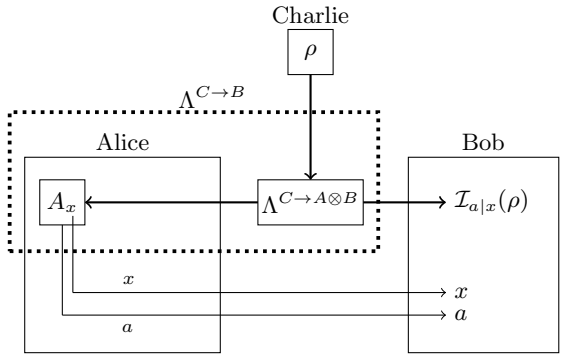
\begin{figure}[t]
\centering
\begin{tikzpicture}

    \node at (0.2, 2) {$\rho$};
    \draw[thick, ->] (0.2,1.7) -- (0.2,0.3);
    \draw (-0.5,-0.3) rectangle (0.9,0.3) node[pos=0.5] {$\Lambda^{C \rightarrow A\otimes B}$};
    \draw[thick, ->] (0.9,0) -- (2,0) node[right] {$\mathcal I_{a|x}(\rho)$};
    \draw[thick, ->] (-0.5,0) -- (-2.8,0);
    \draw (-3.4,-0.3) rectangle (-2.8,0.3) node[pos=0.5] {$A_x$};
    \draw[-]  (-3.1, -0.3) -- (-3.1, -1.5);
    \draw[-]  (-2.96, -0.18) -- (-2.96, -1.2);
    \draw[->] (-3.1, -1.5) -- (2,-1.5) node[pos=0.17, below] {\scriptsize $a$} node[right]{$a$} ;
    \draw[->] (-2.96, -1.2) -- (2,-1.2)  node[pos=0.15, above] {\scriptsize $x$} node[right]{$x$};

    \draw[-] (-0.1,1.7) rectangle (0.5,2.3);
    \draw[-] (-3.6,-2) rectangle (-1,0.6);
    \draw[-] (1.5,-2) rectangle (3.5,0.6);
    \node at (-2.3, 0.8) {Alice};
    \node at (2.5, 0.8) {Bob};
    \node at (.2, 2.5) {Charlie};

    \draw[dotted, line width = 0.5mm] (-3.8,-0.65) rectangle (1.1,1.2);
    \node at (-1.1, 1.4) {$\Lambda^{C \rightarrow B}$};

\end{tikzpicture}
\caption{Channel steering: the setup is similar to the spatial steering scenario, but in the channel case the shared state is prepared by Charlie via the broadcast channel $\Lambda^{C \rightarrow A\otimes B}$. The operations enclosed in the dotted line are then viewed by Bob as instruments which have the total channel $\Lambda^{C \rightarrow B}$. The main difference to spatial steering is that here Bob's task is to build a local (instrument) model for all possible input states, see Eq.~(\ref{instrsteer}).}
\label{Fig2}
\end{figure}

\textit{Channel steering.---} In channel steering \cite{Piani2} one is interested in an assemblage of instruments $\{\mathcal I_{a|x}\}_{a,x}$ instead of states. This assemblage originates from a process where Charlie sends quantum states to Bob through a quantum channel $\Lambda^{C\rightarrow B}$ which possibly entangles some of the states to an environment (Alice) (see Fig.~\ref{Fig2}). The task is to decide if the entanglement between Alice and Bob is strong enough to allow Alice to steer Bob's outputs of the channel. Mathematically this means that one takes a channel extension $\Lambda^{C\rightarrow A\otimes B}$ of the channel $\Lambda^{C\rightarrow B}$ and defines an instrument assemblage through
\begin{align}\label{instrext}
\mathcal I_{a|x}(\rho)=\text{tr}_A[(A_{a|x}\otimes\openone)\Lambda^{C\rightarrow A\otimes B}(\rho)].
\end{align}

Note that here the assemblage is non-signalling by definition. The unsteerability of this instrument assemblage is defined as the existence of a common instrument $\mathcal I_\lambda$ and post-processings $p(a|x,\lambda)$ such that
\begin{align}\label{instrsteer}
\mathcal I_{a|x}=\sum_\lambda p(a|x,\lambda)\mathcal I_\lambda.
\end{align}

Noticing that Eq.~(\ref{instrsteer}) and Eq.~(\ref{instrcomp}) are identical and using a minimal dummy system instead of a generic extension in Eq.~(\ref{instrext}) we arrive to the following Observation.

\begin{observation}\label{ChannelJM}
An instrument assemblage $\{\mathcal I_{a|x}\}_{a,x}$ defined through a minimal dilation $\Lambda^{C\rightarrow A\otimes B}$ is unsteerable if and only if the associated dummy POVMs $\{\tilde A_{a|x}\}_{a,x}$ are jointly measurable.
\end{observation}

There exists a former result on the connection between channel steering and joint measurability \cite{Banik}. This result, however, can be proven to be false (see Appendix B).

With all three steering scenarios being mapped into incompatibility, we are ready to discuss the use of measurement uncertainty relations as steering inequalities.

\textit{Steering inequalities from incompatibility.---} As various joint measurement uncertainty relations have been analytically characterised \cite{Paul, Paul2, YuOh, Liang, Carmeli, Erkka, Kunjwal, Roope3}, our Observation~\ref{steerJM} and Observation~\ref{ChannelJM} open up a possibility to use these relations as steering inequalities for all three scenarios. To give an example of such uncertainty relation, consider the simplest case of two two-valued qubit observables given by
\begin{align}\label{PaulMUR}
A_{\pm|x}=\frac{1}{2}(\openone\pm\vec a_x\cdot\vec\sigma),\quad x=1, \, 2.
\end{align}
These observables are jointly measurable \cite{Paul} if and only if
\begin{align}
\|\vec a_1+\vec a_2\|+\|\vec a_1-\vec a_2\|\leq2.
\end{align}
This inequality is universally applicable to all three steering scenarios and gives an `if and only if' condition for each of them. 
As an example, consider inserting the observables from Eq.~(\ref{PaulMUR}) as the dummy observables to Eq.~(\ref{Bobsolve}). This gives instruments for which channel steering can directly be decided. We are ready to state our next Observation.

\begin{observation}
Joint measurement uncertainty relations can be used as steering inequalities for spatial, temporal and channel steering.
\end{observation}

In the following we further demonstrate the applicability of the technique by first showing an equivalence between temporal and spatial steering and then showing that temporally non-steerable correlations are a proper subset of macrorealistic correlations.

\textit{Equivalence between temporal and spatial steering. --} Applying the Stinespring dilation to a set of non-signalling instruments $\{\mathcal I_{a|x}\}_{a,x}$ shows that the temporal steering scenario they define can be mapped into the spatial steering scenario, see Eq.~(\ref{Stineins}). The question remains which spatial scenarios can be reached by these instruments as the mapping is not in general injective.

To answer this, take a non-signalling state assemblage $\{\rho_{a|x}\}_{a,x}$ with a $d$-dimensional support. Notice that this state assemblage can be prepared through spatial steering using a purification of the total state $\rho_B:=\sum_a\rho_{a|x}$ \cite{Gisin, Hughston, Sainz} $[$see also Eq.~(\ref{AliceBob})$]$. Hence, we need an isometry $V$ which has such purification in its range. One possible choice is the set of Kraus operators $K_k=|k\rangle\langle k|$, where $\{|k\rangle\}_{k=1}^d$ is the eigenbasis of $\rho_B$. Taking the input state $|\psi\rangle:=\sum_{i=1}^d\sqrt{\lambda_i}|i\rangle$, where the numbers $\lambda_i>0$ are the eigenvalues of the state $\rho_B$, and the observables $\tilde A_{a|x}:=\rho_B^{-1/2}\rho_{a|x}^T\rho_B^{-1/2}$, where the transpose is taken in the eigenbasis of $\rho_B$, we get through the minimal dilation of the channel $\Lambda(\rho):=\sum_k K_k\rho K_k^\dagger$ the desired state assemblage
\begin{align}\label{TSSS}
&\mathcal I_{a|x}(|\psi\rangle\langle\psi|)
=\sum_{k,l=1}^d\langle l|\tilde A_{a|x}|k\rangle K_k|\psi\rangle\langle\psi| K_l^\dagger\nonumber\\
&=\sum_{k,l}\tr_A[(\rho_B^{-1/2}\rho_{a|x}^T\rho_B^{-1/2}\otimes\openone)\rho_B^{1/2}|k\rangle\langle l|\rho_B^{1/2}\otimes|k\rangle\langle l|]\nonumber\\
&=\rho_{a|x}.
\end{align}

With this, we are ready to state the next Observation:

\begin{observation}\label{TSSS}
Temporal and spatial steering are fully equivalent problems in that temporal steering can be embedded into the spatial scenario and the two can produce exactly the same assemblages. Moreover, any non-signalling state assemblage on a $d$-level system can be reproduced with non-signalling instruments acting on a $d$-level system.
\end{observation}

The above Observation has two crucial consequences. First, if temporal steering results in a non-signalling assemblage, then it can be realised with non-signalling instruments. Hence, restricting to non-signalling instruments is actually not a restriction at all. Second, Observation~\ref{TSSS} allows one to prove a hierarchy between temporal steering and macrorealistic hidden variable models (see below).

\textit{Temporal steering and macrorealism.---} We now proceed to show that steering has an analogous role in the temporal scenario to that of the spatial case. Namely, whereas spatially non-steerable correlations are a proper subset of local correlations, we show that temporally non-steerable correlations are a proper subset of macrorealistic correlations. To be precise, by correlations we mean here the joint probability distributions and not the correlation functions.

To do so, recall that the probabilities in a sequential measurement scenario (consisting here of two different time steps) are said to have a macrorealistic hidden variable model if they can be written in the form \cite{Kofler}
\begin{align}\label{MRHV}
\text{tr}[\mathcal I_{a|x}(\rho)B_{b|y}]=\sum_\lambda p(\lambda)p(a|x,\lambda)p(b|y,\lambda),
\end{align}
where $p(\cdot), p(\cdot|x,\lambda)$ and $p(\cdot|y,\lambda)$ are probability distributions for all $x,y$ and $\lambda$. Provided that one uses non-signalling instruments, the l.h.s.\ of the above equation can be written in the distributed scenario simply as $\text{tr}[(\tilde A_{a|x}\otimes B_{b|y})V\rho V^\dagger]$. As the non-signalling condition is automatically satisfied for a given total channel, our question boils down to finding an isometry $V$ and a state $\rho$ such that the state $V\rho V^\dagger$ is steerable but local. By defining the Kraus operators
\begin{align}
K_0&=|0\rangle\langle 0|+|1\rangle\langle 1|\\
K_1&=|0\rangle\langle 2|+|1\rangle\langle 3|
\end{align}
we see that the state $\rho:=\lambda |\psi\rangle\langle\psi| + (1-\lambda)\frac{1}{4}\openone_4$, where $|\psi\rangle=\frac{1}{\sqrt 2}(|0\rangle +|3\rangle)$, maps to the isotropic state $V\rho V^\dagger=\lambda|\psi^+\rangle\langle\psi^+|+(1-\lambda)\frac{1}{4}\openone_4$. Isotropic states are steerable but local for projective measurements with $\frac{1}{2}<\lambda\leq\frac{1}{K_G(3)}$, where $K_G(3)$ is a Grothendieck constant and $\frac{1}{K_G(3)}\approx 0.6595$ \cite{Wiseman, Acin}. However, considering only projective measurements does not cover all possible instruments compatible with the total channel but a similar result can be proven for POVMs (see Appendix C). As the (non-tight) inclusion of temporally non-steerable correlations to the set of macrorealistic correlations follows from their definitions, we are ready to write down our last Observation:

\begin{observation}
The set of temporally non-steerable correlations is a proper subset of macrorealistic correlations.
\end{observation}

The above Observation shows that there exists instances of temporal steering where a certain steerable channel-state pair can never lead to non-macrorealistic behaviour, no matter what (non-signalling) measurements (compatible with the channel) are performed on the first party.

\textit{Conclusions. ---} In this work, we have approached spatial, temporal and channel steering through a modified version of the well-known Stinespring dilation. We have demonstrated the power of our approach by showing that incompatibility of quantum measurements is one-to-one connected to quantum steering in all three scenarios. In addition, we have shown how measurement uncertainty relations can be used as universal steering inequalities through this connection.

In contrast to the formerly known connections between spatial steering and joint measurability \cite{Tulio, Roope1, Roope2, Jukka}, the new approach is not limited to incompatibility. Using the Stinespring approach, we have mapped temporal steering into a framework where non-signalling is a built-in state-independent feature. Moreover, we have shown an equivalence between temporal and spatial steering, and shown that temporally unsteerable correlations are a proper subset of non-macrorealistic correlations. For future works it would be interesting to investigate other possible connections between temporal and spatial correlations, e.g., investigate if our approach can be used to translate such concepts as entanglement in a meaningful way to the temporal scenario.

We thank C. Budroni, A. C. S. Costa and M. Banik for stimulating discussions. This work has been supported by ERC (Consolidator Grant 683107/TempoQ), DFG and the Finnish Cultural Foundation.

\section{Appendix}

\subsection{A. Temporal steering}

In a former work \cite{Karthik} the authors have discussed a connection between temporal steering and joint measurability using L\"uders instruments to describe the state update caused by Alice's measurements. The result of the article states that a set of observables is non-jointly measurable if and only if it can be used for temporal steering. Whereas this claim works perfectly for the maximally mixed input state, it is worth noting that, for example, a typical joint measurement scenario with orthogonal noisy qubit observables $A^\eta_{\pm 1|x}:=\frac{1}{2}(\openone\pm\eta\vec{x}\cdot\vec\sigma)$, where $0<\eta\leq1$ is the noise parameter, leads to signalling assemblages with any other input state than the maximally mixed one. Hence, even jointly measurable observables, i.e. $\eta\leq\frac{1}{\sqrt 3}$ \cite{Paul}, can lead to temporal steering in the state-dependent framework providing a counter example for the general claim in \cite{Karthik}.

For scenarios including the maximally mixed input state and L\"uders instruments, one sees that our approach gives the transposed versions of Alice's observables as dummy observables. Hence, one sees that the claims made in \cite{Karthik} for the specific input state and instruments can be reproduced using our method.

\subsection{B. Channel steering}

There exists a former result \cite{Banik} reporting a one-to-one connection between joint measurability of measurements $\{A_{a|x}\}_{a,x}$ on any dilation (or extension) of the total channel and the non-steerability of the instrument assemblage they define. While it is true that compatible measurements will not lead to channel steering no matter which dilation (or extension) is used, the other direction is not true in general. Take, for example, any instrument assemblage $\{\mathcal I_{a|x}\}_{a,x}$ defined through linearly dependent Kraus operators $K_1=\frac{1}{\sqrt 2}U,$ $K_2=\frac{1}{\sqrt 2}U$ of some unitary channel $\Lambda_U(\rho)=U\rho U^\dagger$. The instrument assemblage is given by
\begin{align}
\mathcal I_{a|x}(\rho)=
\frac{1}{2}\sum_{k,l=1}^2\langle\varphi_l|\tilde A_{a|x}|\varphi_{k}\rangle U\rho U^\dagger.
\end{align}
Hence defining $p(a|x,\lambda)=\frac{1}{2}\sum_{k,l}\langle\varphi_l|\tilde A_{a|x}|\varphi_{k}\rangle$ (which is a probability distribution as $\{\tilde A_{a|x}\}_{a,x}$ is a POVM), $\Lambda_\lambda=\Lambda_U$ and the hidden variable space to be trivial, one sees that the setup is unsteerable for compatible as well as for incompatible sets $\{\tilde A_{a|x}\}_{a,x}$ of POVMs.

To see how our result fits to the above example, note that as the {\it minimal} dilation of the channel $\Lambda_U$ is one-dimensional, observables in this space are always jointly measurable and hence the instruments assemblage is non-steerable.

\subsection{C. Steering vs. macrorealism with POVMs on the dilation}

To construct a channel-state pair that admits only local correlations but is still steerable for any possible instrument assemblage we recall that all possible instrument assemblages $\{\mathcal I_{a|x}\}_{a,x}$ compatible with a channel $\Lambda$ are given by the minimal Stinespring dilation:
\begin{align}
\{\mathcal I_{a|x}(\cdot)\}_{a,x}=\{\text{tr}_{\mathcal A}[(\tilde A_{a|x}\otimes\openone)V(\cdot)V^\dagger]|\{\tilde A_{a|x}\}_a\text{ is a POVM}\}.
\end{align}
To provide the desired example, we use a known steerable qutrit-qutrit state which is local for POVMs \cite{Tulio2} as our target state $V\rho V^\dagger$. The state reads
\begin{align}
\tilde\rho:=&\frac{1}{9}\big[a|\varphi^-\rangle\langle\varphi^-|+(3-a)\frac{1}{2}\openone\otimes|2\rangle\langle 2|\\
+&2a|2\rangle\langle 2|\otimes\frac{1}{2}\openone + (6-2a)|22\rangle\langle 22|\big],
\end{align}
where $|\varphi^-\rangle=\frac{1}{\sqrt 2}(|01\rangle-|10\rangle)$, $\openone=|0\rangle\langle 0|+|1\rangle\langle 1|$ and $0<a\leq\frac{3}{2}$. To reach this state we can use a channel acting on $\mathbb C^7$ defined through the Kraus operators
\begin{align}
K_0&=|1\rangle\langle 0|+|2\rangle\langle 2|,\\
K_1&=-|0\rangle\langle 1|+|2\rangle\langle 3|,\\
K_2&=|0\rangle\langle 4|+|1\rangle\langle 5|+|2\rangle\langle 6|.
\end{align}
Now the state
\begin{align}
\rho:=&\frac{1}{9}\big[a|\psi\rangle\langle\psi|+(3-a)\frac{1}{2}(|2\rangle\langle 2|+|3\rangle\langle 3|)\\
+&a(|4\rangle\langle 4|+|5\rangle\langle 5|)+(6-2a)|6\rangle\langle 6|\big],
\end{align}
where $|\psi\rangle=\frac{1}{\sqrt 2}(|0\rangle + |1\rangle)$, maps to the state $\tilde\rho$ on the minimal dilation space, hence completing the example.

\bibliographystyle{apsrev4-1} 

\end{document}